\documentclass[10pt]{article}
\usepackage{latexsym}
\usepackage{graphicx}
\usepackage{times}

\begin{document}

\title{Heuristic explanation of quantum interference
experiments}
\author{Wang Guowen\\College of Physics, Peking University, Beijing, China}
\date{January 25, 2005}
\maketitle

\begin{abstract}
A particle is described as a non-spreading wave packet satisfying
a linear equation within the framework of special relativity.
Young's and other interference experiments are explained with a
hypothesis that there is a coupling interaction between the peaked
and non-peaked pieces of the wave packet. This explanation of the
interference experiments provides a realistic interpretation of
quantum mechanics. The interpretation implies that there is
physical reality of particles and no wave function collapse. It
also implies that neither classical mechanics nor current quantum
mechanics is a complete theory for describing physical reality and
the Bell inequalities are not the proper touchstones for reality
and locality. The problems of the boundary between the macro-world
and micro-world and the de-coherence in the transition region
(meso-world) between the two are discussed. The present
interpretation of quantum mechanics is consistent with the
physical aspects of the Copenhagen interpretation, such as, the
superposition principle, Heisenberg's uncertainty principle and
Born's probability interpretation, but does not favor its
philosophical aspects, such as, non-reality, non-objectivity,
non-causality and the complementary principle.

\end{abstract}
\textbf{Key words:} quantum interference, wave-particle duality,
non-spreading wave packet
\newline
\textbf{PACS:} 03.65.-w

\section{Introduction}
    In 1801 Thomas Young performed a two-pinhole light interference
experiment and demonstrated light to have a wave nature against
the tide of favor for Newton's corpuscular theory of light at that
times [1]. Almost a century after, taking Planck's quantum
hypothesis [2] a big step further, Albert Einstein proposed that
light is made up of particles and successfully explained the
photoelectric effect [3]. His explanation of the effect suggests
that light has wave-particle duality, the central problem of
quantum physics. Later, on the basis of Planck and Einstein's
quantum theories, Louis de Broglie proposed that perhaps matter
also has wave properties [4]. His matter wave hypothesis was
confirmed by Davisson and Germer's electron diffraction experiment
[5]. In addition, in the last century there were two experiments
first demonstrating diffraction and interference of single
particles and illustrating the wave-particle duality: a needle
diffraction experiment with single photons reported in 1909 by
Taylor [6] and an interference experiment with single electrons by
using an electron bi-prism reported in 1989 by Tonomura and
co-workers [7]. However, in the past almost hundred years
physicists and others were unable to find a convincing way to
solve the wave-particle duality problem. It still remains a
baffling and fascinating mystery to us today. As the paradigm
experiment of quantum mechanics, Young's experiment with single
photons is mentioned in more or less details in quantum textbooks.
We learn in them that if the intensity of a beam of
single-frequency light is lowered so much that photons pass one
after the other through two slits at a barrier, after a long time
exposure an interference pattern consisting of tiny discrete spots
emerges on a detector screen some distance behind the barrier. It
implies that the photon is somehow divided into two pieces passing
through both slits simultaneously and interfering with each other.
For the path problem of particles, Heisenberg said: $``$the
$`$path$'$ comes into being only because we observe it$''$ [8].
The Copenhagen interpretation of quantum mechanics asserts that
observation creates reality. Although the assertion has been
accepted readily or reluctantly by many physicists, nevertheless
physicists and students still like to ask: $``$How can a particle
pass through both slits simultaneously and interfere with
itself?$''$ This article attempts to explain quantum interference
experiments, solve the wave-particle duality problem and answer
the question.

\section{Description of particles}
    As well known, de Broglie first introduced a wave group associated with a
particle that its group velocity is, in a first-order
approximation, the velocity of the particle [4]. But the wave
group spreads in time. Later he proposed a pilot-wave theory [9]
and in the mid-fifties of the last century he considered a
particle to be a soliton-like structure satisfying a non-linear
differential equation [10]. Nevertheless he also did not succeed
much in this direction. Perhaps we need to find a way out. Now, as
an exploration of ways, a particle as a wave-particle entity is
described  in the following by means of a non-spreading wave
packet satisfying a linear equation within the framework of
special relativity.

We recall that de Broglie's original theory of matter waves in his
doctoral thesis is essentially relativistic [4]. To gain insight
into the quantum world we still need to resort to the special
theory of relativity. Equivalently to Minkowski's formulation of
special relativity, we consider a four-dimensional system
$(x,y,x,w)$ and assume that a particle moves at the light speed
$c$ in accordance with the following equation:
\begin{equation}
\label{eq1}
c^2t^2=x_0^2+y_0^2+z_0^2+w_0^2, \mbox{ }w_0=c\tau_0
\end{equation}
where $\tau_0$ is the proper time. The variable $w$ is much like
the spatial ones, $x$, $y$ and $z$. From Eq.(1) we can draw a
figure (Fig.1) to show the motion of the particle and illustrate
Einstein's mass-velocity relation.
\begin{figure}[htbp]
\centerline{\includegraphics[width=3.6in,height=2.4in]{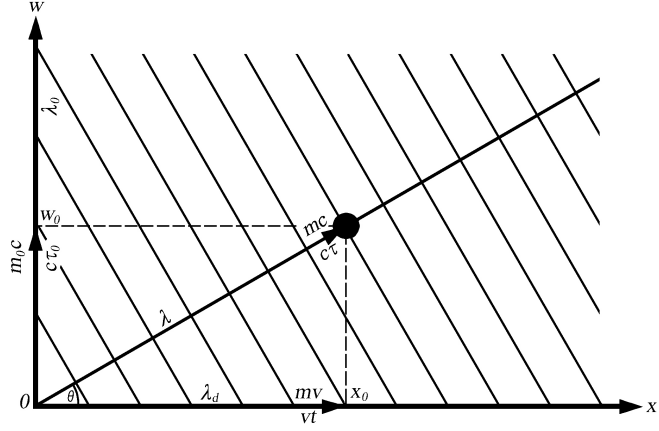}}
\label{Fig.1} \caption{Illustration of Einstein's mass-velocity
relation and graphic display of the Compton wavelength
$\lambda_0$, transformed Compton wavelength $\lambda$ and
relativistic de Broglie wavelength $\lambda_d$.}
\end{figure}
In this figure only the two-dimensional system $(x,w)$ is drawn
for the purpose of this article and $x_0=vt$ is given. Assuming
that the quantities $m_0c$ and $mv$ are the components of
$``$momentum$''$ $mc$ in the $w$ and $x$ directions where $m$ is
the mass of the particle, from this figure we can write
                                                                                                                  the following relations:
\begin{equation}
\label{eq2}
m=\frac{m_0}{\sqrt{1-v^2/c^2}}
\end{equation}
\begin{equation}
\label{eq3}
mc^2=m_0c^2\sqrt{1-v^2/c^2}+mv^2
\end{equation}
Concerning the wave property of the particle, according to de
Broglie [4], we have the Compton wavelength $\lambda_0$,
transformed Compton wavelength $\lambda$ and relativistic de
Broglie wavelength $\lambda_d$:
\begin{equation}
\label{eq4}
\lambda_0=2\pi\hbar/m_0c
\end{equation}
\begin{equation}
\label{eq5}
\lambda=2\pi\hbar/mc
\end{equation}
\begin{equation}
\label{eq6}
\lambda_d=2\pi\hbar/mv
\end{equation}
which are shown in Fig.1.

Now we assume a sharply localized wave packet composed of infinite
harmonic waves propagating at the same phase speed $c$ in the four
dimensions. If the center of the wave packet is at the origin of
the coordinate system when $t$=$0$, as a model, it can be
approximately written as the following Fourier integral associated
with a four-dimensional Dirac delta function:
\begin{equation}
\label{eq7}
B(\textbf{R},t)=\frac{N}{(2\pi)^4}\int{e^{-i(2\pi
vt-\textbf{\scriptsize{K}}\cdot\textbf{\scriptsize{R}})}}
\mbox{d}K_x\mbox{d}K_y\mbox{d}K_z\mbox{d}K_w, \mbox{ }\left|
\textbf{K}\right|=K=2\pi\nu/c
\end{equation}
where $\textbf{R}$=$(x,y,z,w)$ is a four-dimensional position
vector, $\textbf{K}$=$(K_x,K_y,K_z,K_w)$ is a four-dimensional
wave vector and the wave vectors of all the constituent waves are
assumed in the same direction. Since $K=2\pi\nu/c$ has been
assumed, this wave packet satisfies the following linear equation:
\begin{equation}
\label{eq8}
\frac{\partial^2B}{c^2\partial
t^2}-\frac{\partial^2B}{\partial x^2}-\frac{\partial^2B}{\partial
y^2}-\frac{\partial ^2B}{\partial z^2}-\frac{\partial^2B}{\partial
w^2}=0
\end{equation}
Its linearity is in accordance with the linearity of quantum
mechanics, which is essential for the linear superposition
principle of wave.

In order to use the wave packet to describe the particle and make
it being non-spreading, we merely ascribe relativistic energy
$E=mc^2$ and $``$momentum$''$ $P=mc$ of the particle to one
component of the wave packet, called as a characteristic
component. This way to avoid the indeterminacy in momentum and
energy of a wave packet was proposed by H. R. Brown and R. de A.
Martins [11] and G. Wang [12]. Thus, following Planck, Einstein
and de Broglie's quantum theories, assuming that $E=h\nu_c$ and
$P=\hbar K_c$, we write $K$ and $\nu$ as
\begin{equation}
\label{eq9}
K=K_c+K'=\frac{mc}{\hbar}+K'
\end{equation}
\begin{equation}
\label{eq10}
\nu=\nu_c +\nu'=\frac{mc^2}{h}+\frac{cK'}{2\pi}
\end{equation}
Using Eq.(9), Eq.(10) and Eq.(3), from Eq.(7) we write the
following non-spreading wave packet describing the particle:
\begin{eqnarray}
\label{eq11}
B(\textbf{R},t)&=&\frac{N}{(2\pi)^4}e^{-i(2\pi\nu_ct-\textbf{\scriptsize{K}}_c
\cdot\textbf{\scriptsize{R}})} \int
e^{-i(2\pi\nu't-\textbf{\scriptsize{K}}'
\cdot\textbf{\scriptsize{R}})}\mbox{d}{K}'_x\mbox{d}{K}'_y
\mbox{d}{K}'_z\mbox{d}{K}'_w \nonumber\\
&=&\frac{N}{(2\pi)^4}e^{-i[(m_0 c^2\sqrt{1-v^2/c^2}+mv^2)t/\hbar
-\textbf{\scriptsize{K}}_c\cdot\textbf{\scriptsize{R}}]}
\nonumber\\
& & \cdot\int{e^{-i(2\pi\nu't-\textbf{\scriptsize{K}}'
\cdot\textbf{\scriptsize{R}})}}
\mbox{d}{K}'_x\mbox{d}{K}'_y\mbox{d}{K}'_z\mbox{d}{K}'_w
\end{eqnarray}
where $\textbf{K}_c$ is the wave vector of its characteristic
component. The transformed Compton wavelength $\lambda$ equals to
$2\pi/K_c$.

If, for simplicity of treatment, the two-dimensional coordinate
system ($x,w$) is used instead of the four-dimensional system, we
have
\begin{eqnarray}
\label{eq12}
\textbf{K}_c\cdot\textbf{R}
=\frac{mc}{\hbar}(x\cos\theta+w\sin\theta)
=k_cx+\frac{m_0c}{\hbar}w, \nonumber\\
\cos\theta=v/c, \mbox{ }k_c=mc\cos\theta/\hbar
\end{eqnarray}
\begin{equation}
\label{eq13}
\textbf{K}'\cdot\textbf{R}={K}'_x x+{K}'_w w, \mbox{
}{K}'_x ={K}'\cos\theta, \mbox{ }{K}'_w ={K}'\sin\theta
\end{equation}
Using Eq.(12), Eq.(13) and $cK'$=$c(\cos^2\theta+\sin^2\theta
)K'$=$vK'_x+c(1-v^2/c^2)^{1/2}K'_w$, Eq.(11) can be changed into
the following form:
\begin{eqnarray}
\label{eq14}
B(x,w,t)&=&\frac{M}{(2\pi)^2}e^{-i[(m_0
c^2\sqrt{1-v^2/c^2}+mv^2
)t-(\hbar k_cx+m_0 cw)]/\hbar} \nonumber\\
& & \cdot\int{e^{-i[c{K}'t-({K}'_x
x+{K}'_w w)]}} \mbox{d}{K}'_x\mbox{d}{K}'_w \nonumber\\
&=&\frac{M}{(2\pi)^2}e^{-i(vt-x)k_c}\int e^{-i(vt-x){K}'_x}
\mbox{d}{K}'_x \nonumber\\
& & \cdot e^{-i(c\sqrt{1-v^2/c^2}t-w)m_0c/\hbar}\int
e^{-i(c\sqrt{1-v^2/c^2}t-w){K}'_w \mbox{d}{K}'_w}
\nonumber\\
&=&Me^{-i(vt-x)k_c}\delta(vt-x)e^{-i(c\sqrt{1-v^2/c^2}t-w)m_0 c/\hbar} \nonumber\\
& & \cdot\delta(c\sqrt{1-v^2/c^2}t-w)
\end{eqnarray}
where $k_c=mv/\hbar=p/\hbar$ is the component of $\textbf{K}_c$ in
the $x$ direction. From Eq.(14), it is easy to write the
three-dimensional relativistic wave packet that we are really
interested in as follows:
\begin{eqnarray}
\label{eq15}
& & B(\textbf{r},t)=Ce^{-i[(vp+m_0
c^2\sqrt{1-v^2/c^2})t-\textbf{\scriptsize{p}}\cdot
\textbf{\scriptsize{r}}]/\hbar}\delta^3(\textbf{v}t-\textbf{r})=\psi_r(\textbf{r},t)\delta
^3(\textbf{v}t-\textbf{r}), \nonumber\\
& & \psi_r(\textbf{v}t-\textbf{r})=Ce^{-i[(vp+m_0 c^2\sqrt
{1-v^2/c^2})t-\textbf{\scriptsize{p}}\cdot
\textbf{\scriptsize{r}}]/\hbar }, \mbox{ }p=mv, \nonumber\\
& & \delta^3(\textbf{v}t-\textbf{r})
=\delta(v_xt-x)\delta(v_yt-y)\delta (v_zt-z)
\end{eqnarray}
where $\psi_r(\textbf{r},t)$ is the relativistic de Broglie wave
function with the wavelength $\lambda_d=2\pi\hbar/mv$ as shown in
Fig.1. It is worth noting that $m_0c^2(1-\nu^2/c^2)^{1/2}$ may be
considered a $v$-dependent $``$internal energy$''$ and the phase
speed linked to $vp=mv^2$ is exactly the particle speed $v$. When
$v\ll c$, taking the non-relativistic approximation to the first
order in $v$/$c$ and removing the $v$-independent
$\textrm{exp}(-im_0c^2t/\hbar)$, Eq.(15) can be turned into the
non-relativistic wave packet as follows:
\begin{eqnarray}
\label{eq16}
& & B(\textbf{r},t)=Ce^{-i(p^2t/2m_0
-\textbf{\scriptsize{p}}
\cdot\textbf{\scriptsize{r}})/\hbar}\delta^3(\textbf{v}t-\textbf{r})
=\psi(\textbf{r},t)\delta^3(\textbf{v}t-\textbf{r}), \nonumber\\
& & \psi(\textbf{r},t)=Ce^{-i(p^2t/2m_0
-\textbf{\scriptsize{p}}\cdot\textbf{\scriptsize{r}})/\hbar}
\end{eqnarray}
here $p=m_0v$. The $\psi(\textbf{r},t)$, namely the de Broglie
wave function, clearly does not contain information about the
spatial position of the particle and only implies that the
particle has a uniform probability of existing everywhere in the
universe. This explains in a way Born's probability interpretation
of wave function. On the other hand, the function
$\delta^3(\textbf{v}t-\textbf{r})$ describes the trajectory
movement of the peak of the wave packet. The peak can be regarded
as a point-like classical particle. In fact, a real particle has a
finite small size that has been approximated here by infinitesimal
width of the delta function. We are aware that classical mechanics
describes its trajectory movement as well as energy and momentum
that now emerge in the phase of the wave function.

If we regard the spin of the electron, since the Dirac equation
describes a four-component spinor, the spinor is the
characteristic component set of the wave packet describing the
electron. Thus the wave packet would have quadruple components.

Now, we are going to describe a photon. Since it is
$``$massless$''$, $P_w=(K_w)_c\hbar$=0. Regardless of its
polarization or spin, we can describe it as a linear non-spreading
wave packet in which the characteristic component is a scalar
electric wave function. Assuming $\nu$=$kc/{2\pi}$ and denoting
the wave vector of the component by $\textbf{k}_c$, from Eq.(7),
by taking an similar approach used to derive Eq.(15), we write the
wave packet as
\begin{eqnarray}
\label{eq17}
& & B(\textbf{r},t)=\frac{C}{(2\pi)^3}e^{-i(2\pi\nu_c
t-\textbf{\scriptsize{k}}_c\cdot\textbf{\scriptsize{r}})}\int
e^{-i(\textbf{\scriptsize{c}}\cdot
\textbf{\scriptsize{k}}'t-\textbf{\scriptsize{k}}'\cdot
\textbf{\scriptsize{r}})}\mbox{d}{k}'_x \mbox{d}{k}'_y
\mbox{d}{k}'_z \nonumber\\
& & =Ce^{-i(2\pi\nu_c t-\textbf{\scriptsize{k}}_
c\cdot\textbf{\scriptsize{r}})}\delta^3(\textbf{c}t-\textbf{r})
=\varepsilon (\textbf{r},t)\delta^3(\textbf{c}t-\textbf{r}), \nonumber\\
& & \varepsilon(\textbf{r},t)=Ce^{-i(2\pi
\nu_\textrm{\tiny{c}}t-\textbf{\scriptsize{k}}_c
\cdot\textbf{\scriptsize{r}})}, \mbox{ }\left| \textbf{c}
\right|=c
\end{eqnarray}
It is necessary to bear in mind that only the characteristic
component $\varepsilon(\textbf{r},t)$ represents the electric
field associated with the photon and is ascribed the photon energy
$E$=$h\nu_\textrm{\scriptsize{c}}$ and photon momentum $P$=$\hbar
k_\textrm{\scriptsize{c}}$. It would be assumed that a real photon
has a finite small size that has been approximated here by
infinitesimal width of the delta function. Clearly, this model of
the photon reflects Einstein's view that light is described as
$``$consisting of a finite number of energy quanta which are
localized at points in space, which move without dividing, and
which can only be produced and absorbed as complete units$''$ [3].

If we regard the polarization or spin of the photon, since light
wave is a transverse wave of electric and magnetic fields, the set
of the transverse waves is the characteristic component set of the
wave packet describing the photon. Thus the wave packet would have
quadruple components.

\section{Explanation of quantum interference experiments}
The classical explanation of Young's two-slit interference
experiment is based on superposition of waves and the orthodox
explanation on superposition of probability amplitudes and their
collapse during measurement. Now we are going to explain it and
other interference experiments by means of the model of the
non-spreading wave packet. As shown above, the phases of all the
components of the wave packet are the same at the peak and the
distribution of their phases elsewhere makes the resulting
amplitude nearly infinitesimal. From a purely theoretical point of
view, since any piece of the off-peak part is one part of the wave
packet, cutting it away or recombining it will change the path of
the peak and/or the energy of the wave packet. In addition, we are
aware that the concept of a point-like classical particle reflects
the ignorance of the existence of the off-peak part. Thus we think
that this off-peak part possibly plays a dramatic role in the
quantum world.

In order to explain Young's experiment, we consider a barrier in
the $z=0$ plane containing two adjacent long narrow slits parallel
to the y axis labelled by a and b, illuminated by a collimated
beam of linearly polarized single-frequency light. Using Eq.(17),
we assume that one piece
$B_\textrm{\scriptsize{a}}^{(1)}(\textbf{r},t)$ of the wave packet
as photon 1 where the peak of the wave packet locates passes
through slit a and one another
$B_\textrm{\scriptsize{b}}^{(1)}(\textbf{r},t)$ that is non-peaked
passes through slit b simultaneously and then the two are
recombined behind the slits. From the symmetry involved, we also
assume that the peaked piece
$B_\textrm{\scriptsize{b}}^{(2)}(\textbf{r},t)$ of photon 2 passes
through slit b and its non-peaked
$B_\textrm{\scriptsize{a}}^{(2)}(\textbf{r},t)$ passes through
slit a and then they are recombined. To explain the fact that
opening the second slit changes a simple diffraction pattern into
a more complex interference pattern on a detector screen, we are
forced to suppose that there exists a coupling interaction between
the peaked and non-peaked pieces so that the latter shares the
peak and changes the momentum direction of the photon. The above
is a logical answer to the central question: $``$How can a
particle pass through both slits simultaneously and interfere with
itself?$''$ It shows that there is physical reality of particles
and no wave function collapse. Furthermore, since light
interference is correctly understood as a consequence of
Heisenberg's uncertainty principle, this type of coupling
interaction would be the physical origin of the uncertainty
principle.

Let us consider the electric wave function $\varepsilon
(\textbf{r},t)$ in Eq.(17) which is regarded as probability
amplitude. In this two-slit experiment the characteristic
component of the wave packet can be written as $\varepsilon
(x,z,t)$. Now, corresponding with the peaked and non-peaked pieces
of photon 1 and 2,
$B_\textrm{\scriptsize{a}}^{(1)}(\textbf{r},t)$,
$B_\textrm{\scriptsize{b}}^{(1)}(\textbf{r},t)$,
$B_\textrm{\scriptsize{a}}^{(2)}(\textbf{r},t)$ and
$B_\textrm{\scriptsize{b}}^{(2)}(\textbf{r},t)$, we have the
scalar diffracted wave trains, $\varepsilon
_\textrm{\scriptsize{a}}^{(1)}(x,z,t)$,
$\varepsilon_\textrm{\scriptsize{b}}^{(1)}(x,z,t)$,
$\varepsilon_\textrm{\scriptsize{a}}^{(2)}(x,z,t)$ and
$\varepsilon_\textrm{\scriptsize{b}}^{(2)}(x,z,t)$. This
interprets qualitatively the fact that, when a large number of
photons of a collimated beam of single-frequency light land on the
detector screen, an interference pattern emerges.

Similarly, this type of non-spreading wave packet can be used to
explain the experiment with the Mach-Zehnder interferometer with a
photon source, two half-silvered mirrors, two mirrors and two
detectors. If photon 1 is reflected from the first half-silvered
mirror, a coupling interaction occurs when its peaked reflected
piece $B_\textrm{\scriptsize{r}}^{(1)}(\textbf{r},t)$ and
non-peaked transmitted piece
$B_\textrm{\scriptsize{t}}^{(1)}(\textbf{r},t)$ are guided by the
mirrors to the second half-silvered mirror. And, if photon 2
passes through it, another coupling interaction occurs when its
peaked transmitted piece
$B_\textrm{\scriptsize{t}}^{(2)}(\textbf{r},t)$ and non-peaked
reflected piece $B_\textrm{\scriptsize{r}}^{(2)}(\textbf{r},t)$
are guided to the same mirror. Each detector records the number of
both the reflected and transmitted photons from the second
half-silvered mirror. Each number depends on the optical path
lengths of the two arms in the interferometer. In this case, the
result which photon enters which one of the detectors depends on
the result of the coupling interaction between the peaked and
non-peaked pieces of the photon. Meanwhile, if the experimenter
makes a last-moment decision whether to insert the second
half-silvered mirror or not, this decision has no influence on the
past history of the photons. In other words, in this experiment
there is no influence that travels backwards in time. Thus
Wheeler's delayed-choice thought experiment [13] does not
illustrate that there is any influence of the future on the past.

Another typical example is the self-interference of single photons
in a standing wave cavity. Let us consider a cavity with mirrors
on both ends. Assume that an atomic gas is filled inside the
cavity and there exists a standing wave of light with nodes and
antinodes. In this case, according to Born's interpretation of
wave function as probability amplitude, no photon can be found at
the nodes where the value of the wave function is zero. But, this
raises a paradoxical question: $``$How does a photon get through a
node in the standing wave?$''$ Indeed, similar node paradoxes
often arise in atomic and molecular physics. Now we assume that
when the wave packet describing a photon travels back and forth
along the cavity axis, its peaked piece and all forward and
backward travelling non-peaked pieces couple with each other. As a
result, a standing wave as the characteristic component of the
wave packet is formed if the cavity length equals an integer
number of half-wavelengths of the wave. Thus we argue that in
quantum mechanics the predicted zero-probability at the nodes is
no more than reflecting the fact that no quantum effect is caused
there. But this implies by no means that the photon does not pass
through the nodes and is trapped between two adjacent nodes. This
fact of no light-matter interaction with the atoms at the nodes is
just the result of the self-interference of the photon. In other
words, the self-interference of the travelling photon makes itself
inactive at the nodes and maximally active at the antinodes. So we
may still assume a picture underlying quantum mechanics that the
peak of the wave packet has a classical-looking uniform position
probability distribution along the cavity axis. Needless to say,
the picture is pragmatically dispensable in application of quantum
mechanics in which only Born's probability interpretation has an
operational meaning with respect to observations. However the
picture is necessary now for understanding quantum mechanics free
from weirdness. Here we see a good example that the
classical-looking probability of a reality may be quite different
from the probability predicted by quantum mechanics because of its
self-interference. So, if denying the reality and hence the
difference of the two types of probabilities, it is possible to
raise a paradoxical question. As for Bell's inequalities [14],
although the discussion about them and their experimental tests
are out of the scope of this paper, the above explanation also
implies that the experimental violation of the inequalities with
the Bell-type hidden variables which were not mentioned in the EPR
paper [15] can not disprove the EPR argument. This violation
simply signifies that the quantum probability is different from
the classical-like probabilities. The inequalities are thus not
the proper touchstones for reality and locality.

In addition, we can also try to explain the experiments on
interference of independent photon beams at low lever performed by
Pfleegor and Mandel [16]. In the experiments two beams of light
with nearly the same frequency and constant relative initial phase
during the measurement time interval emit from two independent
lasers. To explain the interference pattern observed, we are
forced to suppose that like near-resonance interaction, a coupling
interaction between the peaked piece of a photon in one beam and a
non-peaked piece of photons in the other beam causes interference
when they are joined together. Supposedly, this is as if the
photon fells the non-peaked piece as its own. Thus the
interference of this type is much like the self-interference of
the photon in Young's two-slit experiment. Clearly, the
interference of the independent photon beams also demonstrates
that it is possible to interfere among the different photons in
Young's experiment with a coherent source.

As for electrons, J\"{o}nsson's electron interference experiment
[17] can be explained by using the wave packet expressed by
Eq.(15) and the hypothesis of the coupling interaction in the same
way as the explanation of Young's experiment has been done above.
Thus, the description of the particle as a non-spreading wave
packet satisfying a linear equation can be considered to be
correct and hence provides a realistic interpretation of quantum
mechanics.

This type of wave packet and the coupling interaction might be
used not only to explain interference experiments for subatomic
particles, but also for atoms and molecules, for example, C$_{60}$
molecule of 1 nm in size [18]. They might also be used to explain
the effect of a time gate on energy distribution of photons
passing through it and time coherence effects of photons, for
example, in Einstein's photon-box thought experiment proposed at
the Sixth Solvay Congress in 1930. In addition, they would be
helpful for understanding other quantum phenomena involving
interference, such as, tunnelling of a quantum across a potential
barrier, entanglement of two or more particles and condensation of
identical particles. Clearly, this type of coupling interaction is
local and would be the physical origin of the Bohm non-local
quantum potential [19]. It should be emphasized that the concept
of the coupling interaction is preliminary and open to further
development.

Concerning a macroscopic object, for example, a tiny grain of
sand, roughly speaking, because the outer matter in it, like a
barrier, screens nearly completely the off-peak part of the inner
matter, diffraction and interference do not appear when many
grains of sands pass through two slits. The screen action reduces
with decreasing of the size of the object. This idea eliminates
the theoretically sharp boundary between the macro-world and
micro-world and helps to clarify de-coherence problems in the
transition region (meso-world) between the two.

\section{Discussion and conclusion}
As seen above, a particle can be described as a non-spreading wave
packet satisfying a linear equation within the framework of
special relativity and interference experiments can be explained
with the hypothesis that there is a coupling interaction between
the peaked and non-peaked pieces of the wave packet. The off-peak
part of the wave packet plays a dramatic role in the quantum
world. It seems that this article has solved the wave-particle
duality problem and answered the question $``$How can a particle
pass through both slits simultaneously and interfere with
itself?$''$. This explanation of the interference experiments
provides a realistic interpretation of quantum mechanics. The
interpretation implies that there is physical reality of particles
and no wave function collapse. It also implies that neither
classical mechanics nor current quantum mechanics is a complete
theory for describing physical reality. This conclusion answers
the question raised by Einstein and coworkers [15]: $``$Can
quantum-mechanical description of physical reality be considered
complete?$''$ Thus, the experimental violation of the inequalities
[14] with the Bell-type hidden variables can not disprove the EPR
argument. This violation simply signifies that the quantum
probability is different from the classical-like probability.

The present realistic interpretation of quantum mechanics is
consistent with the physical aspects of the Copenhagen
interpretation, such as, the superposition principle, Heisenberg's
uncertainty principle and Born's probability interpretation, but
does not favor its philosophical aspects, such as, non-reality,
non-objectivity, non-causality and the complementary principle.

\end{document}